\documentclass[preprint,prl,preprintnumbers,amsmath,amssymb]{revtex4}

\usepackage{graphicx}% Include figure files
\usepackage{dcolumn}% Align table columns on decimal point
\usepackage{bm}% bold math

\begin{document}
\title{Hard-sphere behavior in the dynamics of all monoatomic liquids at the de Gennes minimum}

\author{Wouter Montfrooij} \affiliation{Department of Physics and Astronomy, and the Missouri Research Reactor, University of Missouri, Columbia, 65211 MO.}
\begin{abstract}
{We show that the position of the de Gennes minimum in scattering spectra, where the dynamics of liquids shows down, is given by a hard-sphere expression for a range of mono-atomic liquids that crystallize in a close packed structure. This expression relates the position of the minimum to the number density of the liquid, without any adjustable or unknown parameters. We argue that this implies that a liquid can be viewed as a close packed structure of the cages that represent the confinement of atoms by their neighbors. We further show that some metals deviate from this expression, namely those metals that crystallize in a structure that is not close packed. Our expression should prove very useful in identifying what liquids to study in inelastic scattering experiments given that deviations from normal fluid behavior can already be predicted based on the peak position of the static structure factor.}
\end{abstract}
\maketitle

The equations of hydrodynamics tell us how long length scale fluctuations in liquids dacay back to equilibrium. These equations tell us how long a fluctuation lasts for, and how fast they can propagate through the liquid, or how quickly they diffuse away from the point of origin. All this is captured in the Rayleigh-Brillouin triplet that is visible in light scattering experiments\cite{balu}.\\ 

When photons, or neutrons for that matter, are scattered by a liquid, they can create or absorb a sound wave in an inelastic scattering event, or they can probe the diffusion of particles in a quasi-elastic event. The former shows up as two peaks in the scattered intensity located at energy transfers corresponding to $\hbar c_s q$, with $c_s$ the speed of propagation of sound waves and $\hbar q$ the amount of momentum transfered to the liquid\cite{balu}. The width, in energy, of these sound modes increases quadratically with $q$, similar to the $q$-dependence of the  width of the quasi-elastic feature.\\

\begin{figure}        % Additional options h, t, b, p define figure
                            % placement
\begin{center}              % To ensure that the figure is centred horizontally
% Include here some commands to import the figure
\includegraphics*[viewport=10 0 800 710,width=100mm,clip]{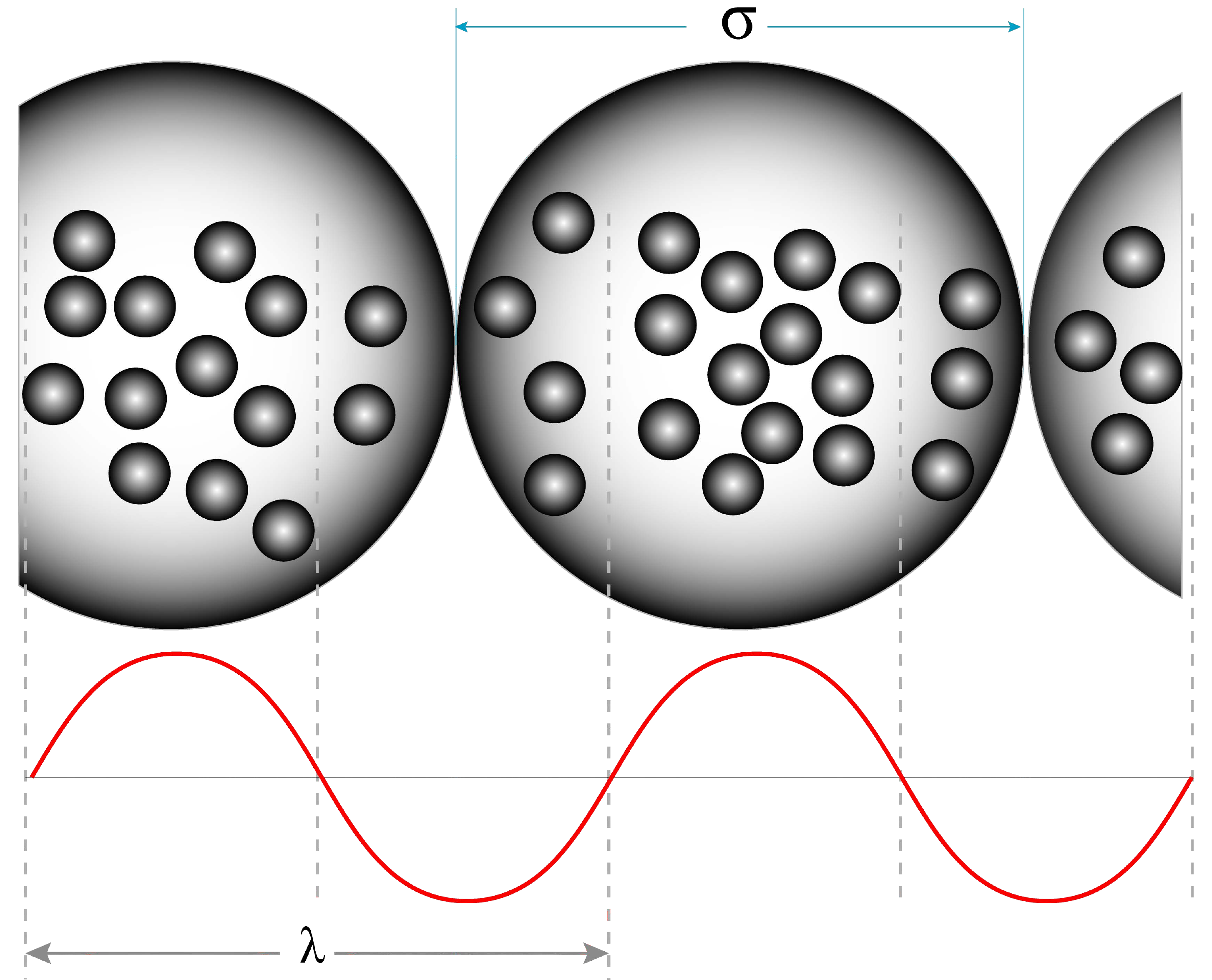}
\end{center}
\caption{Just as a disturbance of a wavelength $\lambda$ spanning quite a few
particles (smaller spheres) can be viewed as a departure from the
average density (9 atoms in between dashed lines in this figure), on a much smaller
scale an atom itself (larger spheres of diameter $\sigma$ a distance $d_{avg}$ apart, shown here for the very high density case $\sigma \approx d_{avg}$) represents a departure from
the average density. Figure rendered by Alexander Schmets.}
\label{densfluc}
\end{figure}

Liquids can also support fluctuations on shorter length scales\cite{gnatzAr,ScopignoReview,wouterbook}, an example of which is shown in fig. \ref{densfluc}. The shorter the length scale of the fluctuation, the more energy it will cost to create. The extreme case would be a fluctuation of wavelength $\lambda$= 2$d_{avg}$ ($d_{avg}$ being the average interatomic separation) where, essentially, we are looking at a fluctuation consisting of a particle next to a hole. Clearly, creating such a hole would be very costly in energy.\\

When the wavelength of the fluctuation is decreased even further, the cost in energy actually goes down again, until there is a minimum in the region when the wavelength corresponds to the interatomic spearation $d_{avg}$. When we probe the liquid on this lengthscale, by transferring an amount of momentum to it given by $q= 2\pi /\lambda= 2\pi /d_{avg}$, we find that fluctuations take longer to decay. This is known as the de Gennes narrowing\cite{gennes}, or structurual slowing down; in scattering experiments we observe that the overall width (in energy) of the spectra reaches a local minimum\cite{balu}. Intuitively this all makes perfect sense: when we try to create a density disturbance of a wavelength corresponding to the natural length scale of the liquid, it would cost a minimum amount of energy. Conversely, such fluctuations would last for a relatively long time since we are actually probing individual particles locked up in a cage of their neighbors.\\

The de Gennes region has received extra scrutiny since this is the region where sound modes can cease to propagate, and where the halfwidths of neutron and X-ray scattering spectra  follow hard-sphere predictions\cite{wouterbook}. In this paper we show that the position (in $q$) where de Gennes narrowing occurs can be accurately predicted for most mono-atomic liquids by a hard-sphere relation between the liquid number density $n$ and the interatomic separation. However, this hard-sphere relation does not imply a hard-sphere potential, rather it simply reflects that the particles in a liquid collide with their neighbors.\\ 

We also show that some mono-atomic liquids, notably Ga\cite{scopignoga} and Hg\cite{badyal}, deviate from our prediction: these are the liquids that crystallize in unusual structures. By comparing the position of the de Gennes narrowing to the density of the liquid, it is possible to predict which liquids will display interesting and unusual dynamics.\\

First we show the validity of our relationship between the position of the de Gennes narrowing and the liquid density, and then we will show how this position can be accurately approximated based on the static structure factor $S(q)$. The latter is useful since it allows for a quick identification of those liquids that merit further, time-consuming inelastic experiments.\\

Thus, we demonstrate the validity of a new relationship between a microscopic quantity that plays a key role in the dynamics of liquids ($q_{min}$) and the macroscopic density. Doing so, we not only arrive at an easy predictor for which liquids are bound to show unusual dynamics, we also find support for a simple picture as to why hard-sphere predictions are able to capture so much of the dynamics of real fluids even when those real fluids do not interact through a potential that (remotely) resembles a hard-sphere interaction.\\

\begin{figure}       % Additional options h, t, b, p define figure
                            % placement
\begin{center}              % To ensure that the figure is centred horizontally
% Include here some commands to import the figure
\includegraphics*[viewport=100 20 650 600,width=100mm,clip]{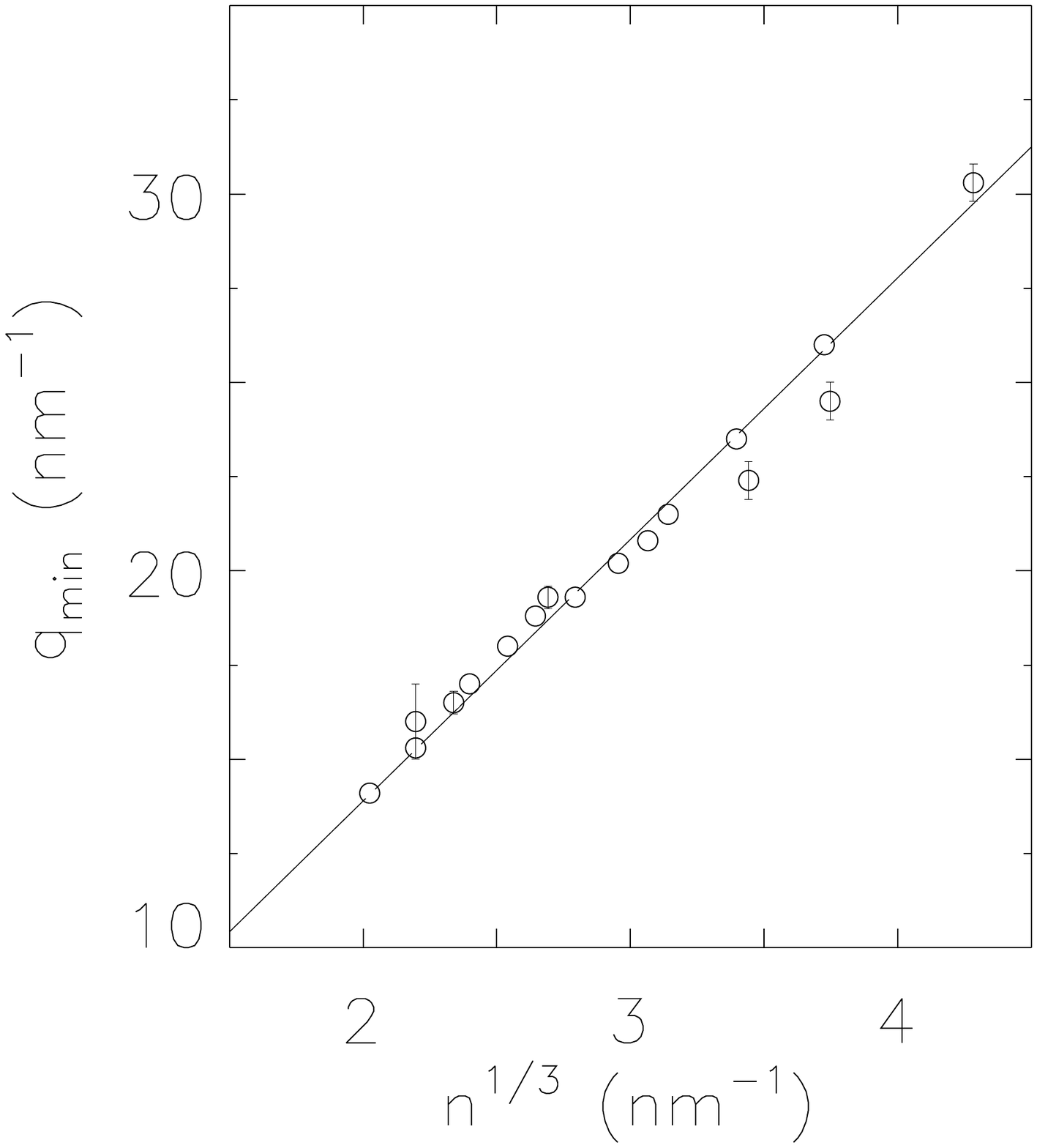}
\end{center}
\caption{The position $q_{min}$ of the de Gennes minimum for a range
of mono-atomic fluids (symbols) compared to the estimate of eqn \ref{main} (solid line). The details are given in Table I. The simple
fluids (inert gases and metals metals) follow the
prediction quite well, whereas Hg and Ga (symbols below the line) clearly
deviate from this prediction. The latter is not entirely unexpected
as these two elements do not condense into a close packed
structure. Typical errorbars on $q_{min}$
are $\pm 0.05$ \AA$^{-1}$ (for details, see Table I), with the errorbars of the outliers shown explicitly.} \label{dylan}
\end{figure}

We show the positions of the de Gennes minimum for a range of fluids in Fig. \ref{dylan}, and list their values in Table I. As can be seen, the positions increase linearly with $n^{1/3}$, as expected: increasing the density by a factor of 8 must lead to a decrease in the average separation by a factor of 2. The straight line that goes through the points, and which captures the exact linear dependence over a range of a factor of 10 in number density $n$,  is given by 
\begin{equation}
q_{min}=\frac{2\pi}{d_{avg}}=\frac{2\pi}{[\pi/(3n\sqrt 2)]^{1/3}}.
\label{main}
\end{equation}
\begin{table}
 {
  \caption{The values for $q_{min}$ for a range of mono-atomic liquids. The liquids were selected from the literature, with the only requirement for inclusion that the position of the minimum could be determined based on the published data. The errorbars on $q_{min}$ are given in brackets. The elements with asterixs$^*$ denote that the minimum was determined based on the minimum of the dispersion relation for the propagating modes, which constitute the entire response in cold helium. Errors in the peak positions of $S(q)$ are roughly 0.5 nm$^{-1}$.}
 \label{equivalent}
 }
 {
  \begin{tabular}{c c c c c c c cl}
    \hline
   &   \\[-2pt]
   element & density & de Gennes minimum & eqn \ref{main}& peak $S(q)$ & peak $S(q)/q^2$ & references \\[3pt]
    & [nm$^{-3}$] &  [nm$^{-1}$]  &  [nm$^{-1}$]  & [nm$^{-1}$]  &  [nm$^{-1}$]  \\[7pt]
      \hline\\[3pt]
cesium & 8.30 &14.1 (3) &14.1&14.3 & 13.9& \onlinecite{bodensteiner}\\[3pt]
rubidium & 10.6 &15.3(5) &  15.3 &15.5&15.2& \onlinecite{rubidium}\\[3pt]
krypton & 10.6 &16.0(10) & 15.3 &18.0 & 16.5 & \onlinecite{egelstaff}\\[3pt]
potassium & 12.8 & 16.5(3) & 16.2 & & &\onlinecite{bermejok}\\[3pt]
krypton & 13.8 &17.0 (10)&16.7 & 17.8 &17.4&\onlinecite{egelstaff}\\[3pt]
helium-3$^*$ & 16.4 &18.0(5)&17.6&&& \onlinecite{albergamohe}\\[3pt]
argon&18.5 & 18.8(3) & 18.4& 19.3 & 18.8 & \onlinecite{vanwellAr}\\[3pt]
argon&19.5 & 19.3(3) & 18.7 & 19.6 & 19.3 & \onlinecite{vanwellAr}\\[3pt]
helium-4$^*$ & 21.8 &19.3(1) &19.4&20.4&19.4& \onlinecite{dietrich,svensson}\\[3pt]
helium-4$^*$ & 25.8 &20.2(2) &20.5&&& \onlinecite{dietrich}\\[3pt]
bismuth & 28.8 & 20.8(5) & 21.3 & 22 & 21 & \onlinecite{bismuth}\\[3pt]
lead & 31.0 & 21.5(5) & 21.8 & 22.1 & 21.5 & \onlinecite{lead}\\[3pt]
%tin$^*$& 32.3 & 21(1) & 22.1 & & & \onlinecite{tin}\\[3pt]
%tin$^*$ & 35.4 &22(2) & 22.8 & 22.6 & 22 &\onlinecite{tin}\\[3pt]
magnesium & 39.2 & 23.5(5) & 23.6 & 24.2& 23.7 &\onlinecite{magnesium}\\[3pt]
mercury&40.8&22.4(5)&23.9&23.3&22.8&\onlinecite{badyal,bafilehg}&\\[3pt]
titanium & 51.7 & 26(2)& 25.9 & &&\onlinecite{titanium}\\[3pt]
gallium &52.6 &24.5 (5) &26.0&25.5 &24.5 &\onlinecite{scopignoga}\\[3pt]
nickel&78.6& 30.3(5) & 29.8 &30.8 &30.5 & \onlinecite{densityni,bermejoni2}\\[3pt]

 \hline\\ %[3pt]
\end{tabular}
  }
\end{table}

\begin{figure}
\begin{center}
\includegraphics*[viewport=25 40 500 185,width=120mm,clip]{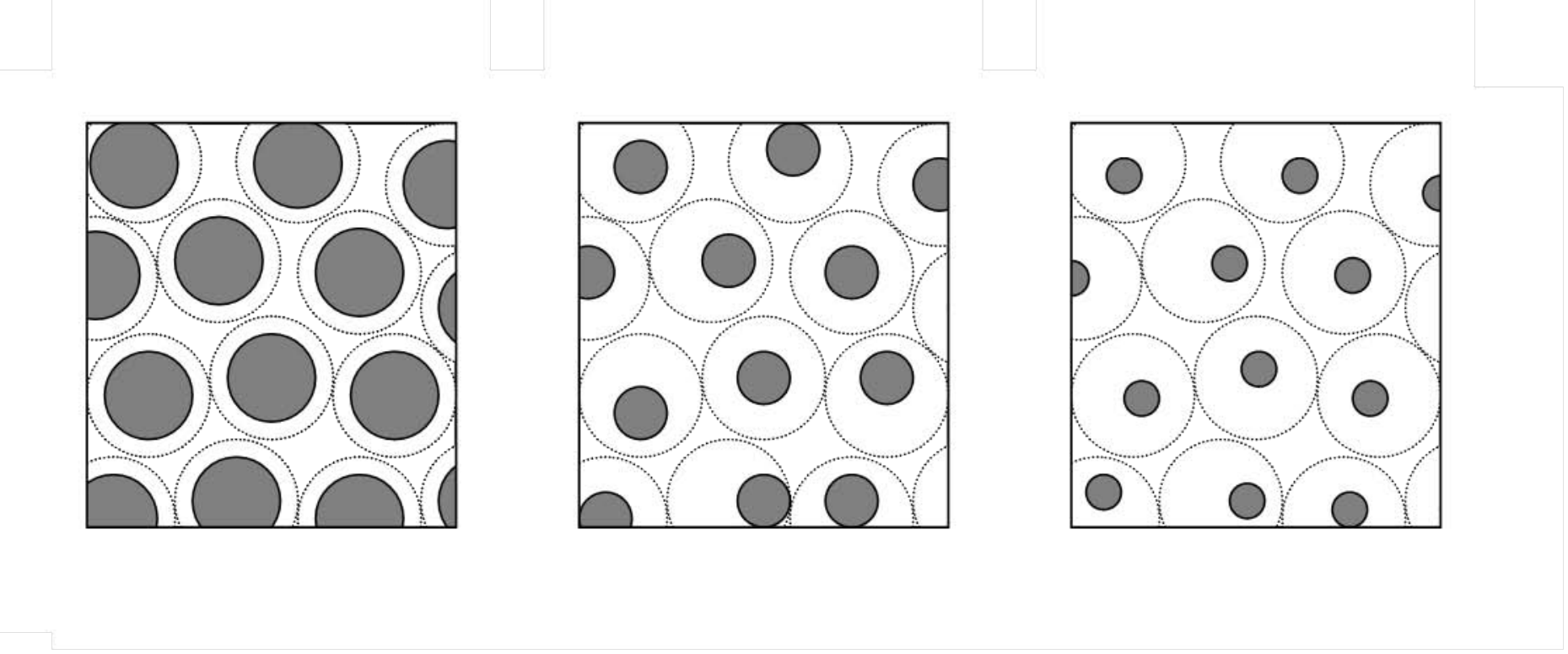}
\end{center}
\caption{The average separation between atoms is, by and large,
given by a close packed structure. The atoms, depicted by the grey
disks, cut out a volume for themselves (shown by the larger,
transparent disks) by colliding with their neighbors and thereby
keeping them at a distance. Since all atoms have very similar
kinetic energies, we can expect the cages that they create to be of
very similar sizes. These cages are stacked in a pattern that
closely resembles a close packed structure. In order to depict a decrease in density in this figure (from left to right), we have reduced the size of the atoms.}
\label{cp}
\end{figure}

The average separation $d_{avg}$ shown in eqn $\ref{main}$ is the separation between particles in a close packed structure of hard-spheres. We show in Fig. \ref{cp} the reason why this expression captures the position of the de Gennes minimum even in cases where the atoms do not touch each other. In a liquid, every atom carves out a little space for itself by colliding with its neighbors. The higher the temperature, the more energetic the collisions and the larger the cage the atom forms for itself. The atoms are essentially locked up in their cages, and even when they do manage to escape their cage, they immediately find themselves in a new cage since they are always surrounded by their neighbors. On short length scales, these cages form a close packed structure, since any other stacking arrangement would correspond to large holes in the liquid, which is, from an energy point of view,  so unfavorable that such defects quickly disappear. This picture also nicely captures why the details of the interatomic interaction do not play much of a role in the exact position of $q_{min}$: cages always form, and their sizes are determined by the amount of kintetic energy available to the collision process.\\

Having validated eqn \ref{main} by comparing the length scale of the slowing down of the dynamics of liquids to the number density, we now turn the equation around in order to use it as a predictor for identifying which liquids would merit further investigation of the dynamics when only static information is known. As can be seen from Table I, the position of the de Gennes minimum and the position of the peak in the static structure factor $S(q)$ are closely related; however, the peak position of $S(q)$ occurs at slightly higher $q$-values and therefore, it cannot be used as a direct measure of the de Gennes minimum.\\

The reason why the maximum of the structure factor does not occur at $2\pi/d_{avg}$ can be seen as follows. If all the atoms would exactly be located at a distance $d_{avg}$ from each other, as is almost the case in a close-packed solid at very low temperatures, then the peak positions would indeed be determined by $2\pi/d_{avg}$. However, atoms stray from their average positions. Picture a pair of atoms in a liquid that are separated by a distance $d_{avg}$ on average, but they can approach each other a little closer or stray away from each other a little bit. Supposing that this movement is symmetric around the equilibrium distance, something which holds true for harmonic motion, then this would cause a shift (in $q$) in peak position. The reason is that in scattering by a liquid that does not possess any preferred direction, the number of $q$-values between $q$ and $q + \Delta q$ is proportional to $q^2\Delta q$. Thus, there are more $q$-values that match the shorter atomic separation $q= 2\pi/r$ than that match the longer atomic separation, even though the changes in atomic separation are completely symmetric around the average separation. The factor $q^2$ pushes the main peak of the strucutre factor out to higher $q$-values, higher than compared to $2\pi/d_{avg}$.\\

Given the preceding paragraph, if we are to determine the average interatomic separation from the static structure factor, then we have to identify the peak position of $S(q)/q^2$. This occurs at lower $q$-values than the position of the maximum of $S(q)$. We show these values in Table I. As can be seen from this table, these values correspond well to the values derived for the position of the de Gennes minimum from inelastic experiments. Thus, one can use measurements of the static structure factor, determine the peak position of $S(q)/q^2$, and compare this position to the prediction of eqn \ref{main}; if the two do not agree, then this liquid would certainly merit an investigation of the dynamics.\\

Our results are actually more useful than appears at first glance. It explains why hard-sphere expressions are so successful at capturing\cite{wouterbook} a large part of the dynamics of liquids whose interatomic interaction potential is anything but hard-sphere like. Every atom carves out a little space for itself by colliding with and pushing other atoms out of the way. These spaces, or cages, naturally arrange themselves in a close packed structure, or at least, or close approximation thereof. When we probe the dynamics on different length scales, we see the motion (wiggling and sliding) of these cages superimposed on the motion of individual atoms. And last but not least, eqn \ref{main} is actually the only equation that we know of that makes a prediction about the microscopic dynamics based on a single  macroscopic property without having to resort to any adjustable parameters.\\

We expect that the motion and rearrangment of the cages is very similar between all liquids since it is a morphological problem rather than an interaction problem. As such, this part of the dynamics should be well captured by any type of interaction potential, such as a hard-sphere interaction. In fact, a hard-sphere interaction potential should exactly capture the dynamics of the cages since this particular interaction potential only springs into action at contact. Given that these moments of contact are actually what creates the cages in the first place, we can view the dynamics of a hard-sphere liquid as being representative of the motion of the cages. By the same token, the differences between real liquids and hard-sphere liquids will be small when probed on length scales outside of the hydrodynamic regime, with the differences only reflecting the effects of the interaction potential during a collision. The validity of eqn \ref{main} supports this assertion. Finally, while we have only applied eqn \ref{main} to monoatomic fluids, it would be an equally useful tool for molecular liquids as it will shed light on how strongly the dynamics are influenced by the cage forming tendencies of molecules.\\

\end{document}